\documentclass[sigconf]{aamas} 

\usepackage{balance} 
\usepackage{subfig}
\usepackage{blindtext}  
\usepackage{multirow}
\usepackage[ruled,linesnumbered]{algorithm2e} 

\usepackage{xcolor}

\newcommand{\algo}{\textsc{hammer}}

\setcopyright{none}

\acmSubmissionID{242} 

\title[]{\algo: Multi-Level Coordination of Reinforcement Learning Agents via Learned Messaging}

 \author{Nikunj Gupta}
 \affiliation{
 	\institution{IIIT-Bangalore}
 	}
 \email{Nikunj.Gupta@iiitb.org}

 \author{G. Srinivasaraghavan}
 \affiliation{
 	\institution{IIIT-Bangalore}
 	}
 \email{gsr@iiitb.ac.in} 

 \author{Swarup Kumar Mohalik}
 \affiliation{
 	\institution{Ericsson Research} 
 	}
 \email{swarup.kumar.mohalik@ericsson.com} 

 \author{Nishant Kumar} 
 \affiliation{
 	\institution{IIT-BHU} 
 }
 \email{nishantkr.ece18@itbhu.ac.in} 

 \author{Matthew E. Taylor} 
 \affiliation{
 	\institution{University of Alberta \&
 	Alberta Machine Intelligence Institute (Amii)}
 }
 \email{matthew.e.taylor@ualberta.ca}

\begin{abstract} 
Cooperative multi-agent reinforcement learning (MARL) has achieved significant results, most notably by leveraging the representation-learning abilities of deep neural networks. However, large centralized approaches quickly become infeasible as the number of agents scale, and fully decentralized approaches can miss important opportunities for information sharing and coordination. Furthermore, not all agents are equal --- in some cases, individual agents may not even have the ability to send communication to other agents or explicitly model other agents. This paper considers the case where there is a single, powerful, \emph{central agent} that can observe the entire observation space, and there are multiple, low-powered \emph{local agents} that can only receive local observations and are not able to communicate with each other. The central agent's job is to learn what message needs to be sent to different local agents based on the global observations, not by centrally solving the entire problem and sending action commands, but by determining what additional information an individual agent should receive so that it can make a better decision. In this work we present our MARL algorithm \algo, describe where it would be most applicable, and implement it in the cooperative navigation and multi-agent walker domains.  Empirical results show that 1) learned communication does indeed improve system performance, 
2) results generalize to heterogeneous local agents, and 3) results generalize to different reward structures.
\end{abstract}


\newcommand{\BibTeX}{\rm B\kern-.05em{\sc i\kern-.025em b}\kern-.08em\TeX}

\begin{document}


\pagestyle{fancy}
\fancyhead{}


\maketitle 


%
%
%
%
%

\section{Introduction} 
\label{sec:introduction}
The field of multi-agent reinforcement learning (MARL) combines ideas from single-agent reinforcement learning (SARL), game theory, and multi-agent systems. Cooperative MARL calls for simultaneous learning and interaction of multiple agents in the same environment to achieve shared goals. Applications like distributed logistics \cite{ying2005multi}, package delivery \cite{seuken2012improved}, and disaster rescue \cite{parker2016exploiting} are domains that can  be modeled naturally using this framework. However, even cooperative MARL suffers from several complications inherent to multi-agent systems, including non-stationarity ~\cite{busoniu2008comprehensive}, a potential need for coordination ~\cite{busoniu2008comprehensive}, the curse of dimensionality ~\cite{shoham2007if}, and global exploration ~\cite{matignon2012independent}. 

Multi-agent reasoning has been extensively studied in MARL ~\cite{matignon2012independent, panait2005cooperative} in both centralized and decentralized settings. While very small systems could be completely centralized, decentralized implementation becomes indispensable as the number of agents increase, and to cope with the exponential growth in the joint observation and action spaces. However, it often suffers from synchronization issues \cite{matignon2012independent} and complex teammate modeling \cite{albrecht2018autonomous}. Moreover, independent learners may have to optimize their own, or the global, reward from only local, private observations ~\cite{tan1993multi}. In contrast, centralized approaches can leverage global information and mitigate non-stationarity through full awareness of all teammates.

In MARL, communication has been shown to be an important aspect, especially in tasks requiring coordination. For instance, agents tend to locate each other or the landmarks more easily using shared information in navigation tasks \cite{fox2000probabilistic}. Communication can also influence the final outcomes in group strategy coordination \cite{wunder2009communication, ito2010innovations}. There have been significant achievements using explicit communication in video games like StarCraft II \cite{peng2017multiagent} 
as well as in mobile robotic teams  ~\cite{matignon2012coordinated}, smart-grid control \cite{pipattanasomporn2009multi}, and autonomous vehicles \cite{cao2012overview}. Communication can be in the form of sharing experiences among the agents ~\cite{zaiem2019learning}, sharing low-level information like gradient updates via communication channels \cite{foerster2016learning} or sometimes directly advising appropriate actions using a pretrained agent (teacher) \cite{2014connectionscience-taylor} or even learning teachers \cite{omidshafiei2019learning,JAAMAS20-Leno}.  

Inspired by the advantages of centralized learning and communication for synchronization we propose multi-level coordination among intelligent agents via messages learned by a separate agent to ease the localized learning of task-related policies. We propose a single \emph{central agent} designed to learn high-level messages based on complete knowledge of all the local agents in the environment. These messages are communicated to the \emph{independent learners} who are free to use or discard them while learning local policies to achieve a set of shared goals. By introducing centralization in this manner, the supplemental agent can play the role of a \emph{facilitator} of learning for the rest of the team. Furthermore, 
independent learners need not 
be as powerful if they must train to communicate or model other agents alongside learning task-specific policies.

A hierarchical approach to MARL is not new --- we will contrast with other existing methods in Section~\ref{sec:related}. However, the main insight of our algorithm is to learn to communicate relevant pieces of information from a global perspective to help agents with limited capabilities improve their performance. 
Potential applications include autonomous warehouse management ~\cite{enright2011optimization} and traffic light control ~\cite{liu2017cooperative}, where there can be a centralized monitor. After we introduce our algorithm, \algo, we will show results in two very different simulated environments to showcase its versatility. The multi-agent cooperative navigation lets agents learn in a continuous state space with discrete actions and global team rewards. Stanford's multi-agent walker environment has a continuous action space and agents can receive only local rewards.  


The main contributions of this paper are to explain a novel and important setting that combines agents with different abilities and knowledge (Section~\ref{setting}), introduce the \algo\ algorithm that addresses this setting (Section~\ref{approach}), and then empirically demonstrate that \algo\ can make significant improvements in learning decision policies for agents (Section~\ref{results}) in two multi-agent domains.

\section{Background and Related Work} 
\label{sec:related}

This section will provide a summary of background concepts necessary to understand the paper and a selection of related work. 

\subsection{Proximal Policy Optimization} 

A popular choice for solving RL tasks is to use policy gradients, where the parameters of the policy $\theta$ are directly updated to maximize an objective $\mathbf{J}$($\theta$) by moving in the direction of $\nabla \mathbf{J}$($\theta$). 
In our work, we make use of Proximal Policy Optimization (PPO) \cite{schulman2017proximal}, which reduces challenges like exhibiting high variance gradient estimates, being sensitive to the selection of step-size, progressing slowly, or encountering catastrophic drops in performance. Moreover, it is relatively easy to implement. 
Its objective function, well-suited for updates using stochastic gradient descent, can be defined as follows: 
$L^{CLIP}(\theta) = \mathbf{E}_t \left[min(r_t(\theta)A_t, clip(r_t(\theta), 1 - \epsilon, 1 + \epsilon)A_t)\right], $
where $r_t$ is the ratio of probability under the new and old policies, $A_t$ is the estimated advantage at time t, and $\epsilon$ is a hyperparameter. 


\subsection{Multi-Agent Reinforcement Learning}
Single-agent reinforcement learning can be generalized to competitive, cooperative, or mixed multi-agent settings. We focus on the fully cooperative multi-agent setting, which can described as a generalization of MDPs to stochastic games (SGs). A SG for $n$ local agents, and an additional centralized agent in our case, can be defined as the tuple $\langle$ $S$, $U$, $O_1$, $\dots$, $O_n$, $A_1$, $\dots$, $A_n$, $P$, $R$, $\gamma$ $\rangle$, where $S$ is the set of all configurations of environment states for all agents, U is the set of all actions for the central agent, $O_1, \dots, O_n$ represent the observations of each local agent, $A_1, \dots, A_n$ correspond to the set of actions available to each local agent and $P$ is the state transition function. $\gamma$ is the discount factor. In case all the agents have the same common goal (i.e., they aim to maximize the same expected sum) the SG becomes fully cooperative. 

Multi-agent state transitions are a result of the joint action of all the agents U $\times$ $A_1 \times \dots \times A_n$. The policies form a joint policy h: S $\times$ U $\times$ \textbf{A}. There can be two possible reward structures for the agents.  First, they could share a common team reward signal, $R: S \times A \rightarrow \Re$, defined as a function of the state s $\in$ S and the agents joint action A: $A_1$ $\times$ ... $\times$ $A_n$. In the case of such shared rewards, agents aim to directly maximize the returns for the team. Second, each agent could receive its own reward $R_i: O_i \times A_i \rightarrow \Re$. A localized reward structure means that agents maximize their own individual expected discounted return r = $\sum_{t=0}^{\infty} \gamma^{t} r^{t}$. 


\subsection{Relevant Prior Work}
%
Recent works in both SARL and MARL have employed deep learning methods to tackle the high dimensionality of the observation and action spaces \cite{mnih2013playing, lillicrap2015continuous, tampuu2017multiagent, foerster2017counterfactual, peng2017multiagent}. 

Several works in the past have taken advantage of hierarchical approaches to MARL. The MAXQ algorithm was designed to provide for a hierarchical break-down of the reinforcement learning problem by decomposing the value function for the main problem into a set of value functions for the sub-problems \cite{dietterich2000hierarchical}. Tang et al. \cite{tang2018hierarchical} used temporal abstraction to let agents learn high-level coordination and independent skills at different temporal scales together. Kumar et al. \cite{kumar2017federated} present another framework benefiting from temporal abstractions to achieve coordination among agents with reduced communication complexities. These works show positive results from combining centralized and decentralized approaches in different manners and are therefore closely related to our work. Vezhnevets et al. \cite{vezhnevets2017feudal} introduce Feudal networks in hierarchical reinforcement learning and employ a Manager-Worker framework. However, there are some key differences. In their case, the manager directly interacts with the environment to receive the team reward and accordingly distributes it among the workers (analogous to setting their goals), whereas in our work the central agent interacts indirectly and receives the same reward as the local agents. Here, the central agent is only allowed to influence the actions of independent learners rather than set their goals explicitly. One further distinction is that they partly pretrain their workers before introducing the manager into the scene. In contrast, our results show that when the central agent and the independent learners simultaneously learn, they achieve better performance. 

Some works have developed architectures that use centralized learning but ensure decentralized execution. COMA \cite{foerster2017counterfactual} used a centralized critic to estimate the Q-function along with a counterfactual advantage for the decentralized actors in MARL. The VDN \cite{sunehag2017value} architecture trained individual agents by learning to decompose the team value functions into agent-wise value functions in a centralized manner. QMIX \cite{rashid2018qmix} employs a mixing network to factor the joint action-value into a monotonic non-linear combination of individual value functions for each agent. Another work, MADDPG \cite{lowe2017multi}, extends deep deterministic policy gradients (DDPG) \cite{lillicrap2015continuous} to MARL. They learn a centralized critic for each agent and continuous policies for the actors and allow explicit communication among agents. Even though the prior research works mentioned here address a similar setting and allow for using extra globally accessible information like in our work, they mainly aim at decentralized execution which applies to domains where global view is unavailable. In contrast, we target domains where a global view is accessible to a single agent, even during execution. 

There has been considerable progress in learning by communication in cooperative settings involving partially observable environments. Reinforced Inter-Agent Learning (RIAL) and Differentiable Inter-Agent Learning (DIAL) ~\cite{foerster2016learning} use neural networks to output communication messages in addition to the agent's Q-values. RIAL used a  shared network to learn a single policy whereas DIAL used gradient sharing during learning and communication actions during execution. Both methods use discrete communication channels. On the other hand, CommNet ~\cite{sukhbaatar2016learning}, used continuous vectors, enabled multiple communication cycles per time step and the agents were allowed to freely enter and exit the environment. Lazaridou et al. ~\cite{lazaridou2016multi} and Mordatch et al. ~\cite{mordatch2018emergence} trained the agents to develop an emergent language for communication. Furthermore, standard techniques used in deep learning, such as dropout ~\cite{srivastava2014dropout} have inspired works where messages of other agents are dropped out during learning to work well even in conditions with only limited communication feasible ~\cite{kim2019message}. However, in all these works, the goal is to learn inter-agent communication alongside local policies that suffer from the bottleneck of simultaneously achieving effective communication and global collaboration ~\cite{sheng2020learning}. They also face difficulty in extracting essential and high-quality information for exchange among agents ~\cite{sheng2020learning}. Further, unlike \algo, these works expect that more sophisticated agents are available in the environment in terms of communication capabilities or the ability to run complex algorithms to model other agents present.

One of the popular ways to carry out independent learning is by emergent behaviours ~\cite{leibo2017multi, leibo2018malthusian, tampuu2017multiagent}, where each agent learns its own private policy and assumes all other agents to be a part of the environment. This method disregards the underlying assumptions of single-agent reinforcement learning, particularly the Markov property. Although this may achieve good results ~\cite{matignon2012independent}, it can also fail due to non-stationarity ~\cite{tuyls2012multiagent, laurent2011world}. Self-play can be a useful concept in such cases ~\cite{silver2016mastering, tesauro1995temporal, bowling2015heads}, but it is still susceptible to failures through the loss of past knowledge ~\cite{leibo2019autocurricula, samothrakis2012coevolving, lanctot2017unified}. Gupta et al. ~\cite{gupta2017cooperative} extend three SARL algorithms, Deep Q Network (DQN) ~\cite{mnih2015human}, Deep Deterministic Policy Gradient (DDPG) ~\cite{lillicrap2015continuous}, and Trust Region Policy Optimization (TRPO) ~\cite{schulman2015trust}, to cooperative multi-agent settings. 


\section{Setting}
\label{setting}

\begin{figure}[tp]
  	\centering
  	\includegraphics[width=0.75\linewidth]{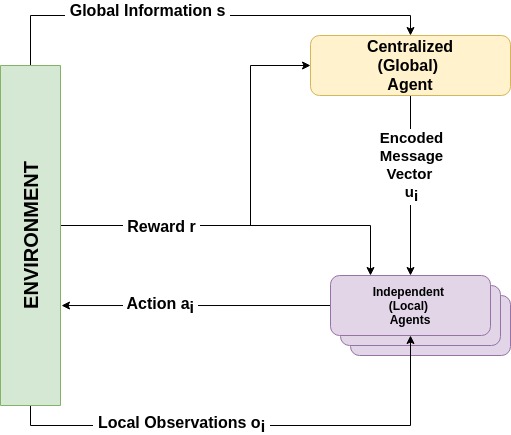}
  	\caption{Our cooperative MARL setting: a single global agent sends messages to help multiple independent local agents act in an environment} 
  	\label{fig:setting}
  \end{figure}

This section details a novel setting in which multiple agents --- the central agent and the local agents --- with different capabilities and knowledge are combined (see Figure~\ref{fig:setting}). The \algo\ algorithm is designed for this type of cooperative multi-agent environment.

Consider a warehouse setting where lots of small, simple, robots fetch and stock items on shelves, as well as bring them to packing stations. If the local agents could communicate among themselves, they could run a distributed reasoning algorithm, but this would require more sophisticated robots and algorithms. Or, if the observations and actions could be centralized, one very powerful agent 
could determine the joint action of all the agents, but this would not scale well and would require a very powerful agent (to address an exponential growth in the observation and actions spaces with the number of agents). Section~\ref{results} will make this more concrete with two multi-agent tasks. Now, assume there is an additional central agent in the team, that has a global perspective, unlike the local agents, who receive only local observations. Further, the central agent is more powerful --- not only does it have access to more information, but it can also communicate messages to all local agents. The local agents can only transmit their observations and actions to the central agent and receive messages --- local agents rely on the communicated messages to know about other agents in the environment. The central agent must learn to encapsulate available information into small messages to facilitate local agents. 

Having described an overview of the setting, we can take a closer look at the inputs, outputs, and roles of the agents in the heterogeneous system as described in Figure ~\ref{fig:setting}. The centralized agent receives a global observation $s \in S$ on every time step and outputs a unique message (equivalently, executes an action), $u_i \in U$, to each of the local agents, where $i$ is the agent identifier. Its global observation s is the union of all the local observations $o_i \in O_i$ and actions of the independent learners $a_i \in A_i$ --- can either be obtained from the environment or transmitted to it by the local agents at every time step. $u_i$ encodes a message vector that a local agent can use to make better decisions. Local agents receive a partial observation $o_i$, and a private message $u_i$ from the central agent. Based on $o_i$ and $u_i$, at each time step, all $n$ local agents will choose their actions simultaneously, forming a joint action ($A_1 \times \dots \times A_n$) and causing a change in the environment state according to the state transition function $P$.  

Upon changing the dynamics of the environment, a reward $r \in R$ --- which could be team-based or localized --- is sent back to the local agents, using which they must learn how to act. If the messages from the central agent were not useful, local agents could learn to ignore them. Every time the central agent communicates a message $u_i$ to a local agent, it learns from the same reward as is obtained by that local agent on performing an action $a_i$ in the environment. In other words, the central agent does not directly interact with the environment to get feedback, instead, it learns to output useful messages by looking at how the independent agents performed in the environment using the messages it communicated. In domains with localized rewards for agents, the central agent gets a tangible feedback for sent messages, whereas, in the case of team rewards, it needs to learn using comparatively abstract feedback. In Section~\ref{results} we show that \algo\ generalizes to both the reward structures. 

\section{The \algo\ Algorithm} 
\label{approach}
This section introduces \algo, the  \emph{Heterogeneous Agents Mastering Messaging to Enhance Reinforcement learning} algorithm, designed for the cooperative MARL setting discussed above. 

There are multiple local and independent agents in an environment that are given tasks. Depending on the domain, they may take discrete or continuous actions to interact with the environment. In \algo, we introduce a single, relatively powerful central agent into the environment, capable of 1) obtaining a global view of all the other agents present in an environment, and 2) transmitting messages to them. It learns a separate policy and aims to support the local team by communicating short messages to them. It is designed to use both a global or local reward structure. As a result, local agents' private observations will now have additional messages sent to them by the central agent and they can choose to use or discard this information while learning their policies. 

\begin{algorithm} [tp]
	\caption{\algo} 
	\label{algorithm}
	\SetAlgoLined 
	\textbf{Initialize} Actor-Critic Network for independent agents (shared network) (IA), Actor-Critic Network for the central agent (CA) or a multi-layered perceptron if gradients from IA are backpropagated to CA, and two experience replay memory buffers (B and B')\; 
	\For{episode e = 1 to TOTAL\_EPISODES}{ 
		Fetch combined initial random observations $s = [o_1, ..., o_i]$ from environment ($o_i$ is agent i's local observation)\; \label{line:state}
		\textit{Input:} Concat ($s = [o_1, ..., o_i]$) $\rightarrow$ CA \; 
		\For{time step t = 1 to TOTAL\_STEPS}{
			\For{each agent $n_i$}{
				\textit{Output}: message vector $u_i$ $\leftarrow$ CA, for agent $n_i$\; \label{message}
				\textit Pass each message vector $u_i$ through a regularization unit \; \label{dru} 
				\textit{Input:} Concat ($o_i \in s$, $u_i$) $\rightarrow$ IA \; \label{learner_input}
				\textit{Output}: local action $a_i$ $\leftarrow$ IA \; \label{local_action}
			} 
			Perform sampled actions in environment and get next set of observations s' and rewards $r_i$ for each agent \; \label{dynamics}
			Add experiences in B and B' for CA and IA respectively\; 
			\If{update interval reached}{
				Sample random minibatch b $\in$ B and b' $\in$ B'\; \label{updates-1}
				Train IA on b' using stochastic policy gradients \; 
				Train CA on b or directly by backpropagating gradients from IA \; \label{updates-2} 
			}
		}
	}  
\end{algorithm}

As described by Algorithm \ref{algorithm}, in every iteration, the centralized agent receives the union of private observations of all local agents (line~\ref{line:state}). It encodes its input and outputs an individual message vector for each agent (line~\ref{message}). These messages from the central agent are sent to the independent learners, augmenting their private partial observations obtained from the environment (line~\ref{learner_input}). Then, they output an action (line~\ref{local_action}) affecting the environment (line~\ref{dynamics}). Reward for the joint action performed in the environment is returned (line~\ref{dynamics}) and is utilized as feedback by all the agents to adjust their parameters and learn private policies (lines~\ref{updates-1}---\ref{updates-2}).

There are multiple techniques for training the central agent to learn how to communicate. One strategy could be to employ any RL algorithm to train \algo's central agent to learn its policy, and use the local agents' reward as a gradient signal. A second strategy would be to push gradients from the local agent's network to \algo's central agent network, by directly connecting the latter's outputted communication actions to the input of the local agent's network. This strategy is inspired from a number of other relevant works \cite{sukhbaatar2016learning, NIPS2016_6042}. Another strategy could be to 
allow messages $m_i$ output by the central agent to first be processed by a regularisation unit, like RU($m_i$) = Logistic($\mathbf{N}$ ($m_i$, $\sigma$)), where $\sigma$ is the standard deviation of the noise added to the channel, and then be passed to the local agent's network \cite{NIPS2016_6042, hinton2011discovering, courbariaux2016binarized}. In Section \ref{results}, we compare our results for all of 
these strategies.

The independent learners follow the formerly proposed idea of allowing all of them to share the parameters of a single policy, 
hence enabling a single policy to be learned with experiences from all the agents simultaneously \cite{gupta2017cooperative, NIPS2016_6042}. This still ensures different behavior among agents because each of them receive different observations. Parameter sharing has been successfully applied in several other multi-agent deep reinforcement learning settings ~\cite{sukhbaatar2016learning, peng2017multiagent, foerster2017stabilising, sunehag2018value, rashid2018qmix}. Learning this way makes it centralized, however, execution can be decentralized by making a copy of the policy and using them individually in each of the agents. Furthermore, in cooperative multi-agent settings, concurrent learning often does not scale up to the number of agents which can make the environment dynamics non-stationary, and difficult to learn \cite{gupta2017cooperative}. Hence, in \algo, a single policy is learned for the local agents via the centralized learning, decentralized execution setting. 

Note that while we focus on having a shared network for independent agents and using PPO methods for policy learning, \algo\ can be implemented with other MADRL algorithms. Moreover, even though we test with a single central agent in this paper, it is entirely possible that multiple central agents could better assist independent learners. This is left to future works. 


\section{Tasks and Implementation Details} 
\label{tasks}
This section details the two multi-agent environments used to evaluate \algo. In addition to releasing our code after acceptance, we fully detail our approach so that results are replicable. 

\subsection{Cooperative Navigation} 
\begin{figure}[t]
	\centering
	\includegraphics[width=0.50\linewidth]{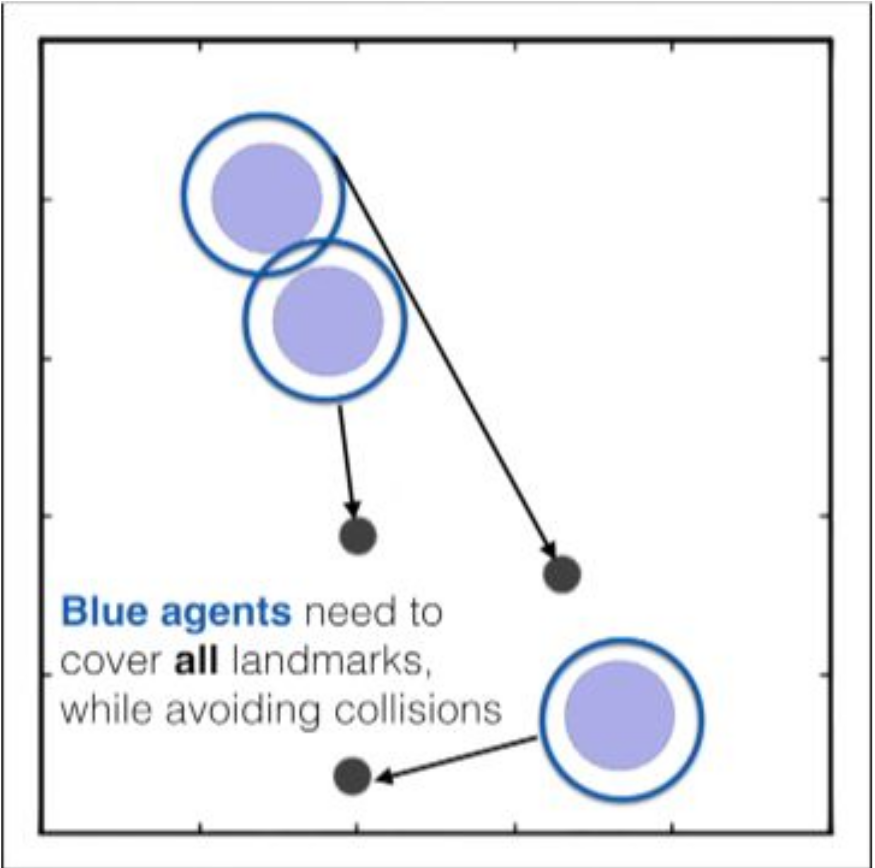} 
	\caption{Cooperative Navigation 
}
	\label{fig:cooperative_navigation}
	\Description{Test bed for the proposed architecture.}
\end{figure}

Cooperative navigation is one of the Multi-Agent Particle Environments~\cite{lowe2017multi}. It is a two-dimensional cooperative multi-agent task with a continuous observation space and a discrete action space consisting of $n$ movable agents and $n$ fixed landmarks. Figure ~\ref{fig:cooperative_navigation} shows the case where $n=3$. Agents occupy physical space (i.e., are not point masses), perform physical actions in the environment, and have the same action and observation spaces. The agents must learn to cover all the landmarks while avoiding collisions, without explicit communication. The global reward signal, seen by all agents, is based on the proximity of any agent to each landmark and the agents are penalized upon colliding with each other. The team reward can be defined by the equation: 
	\[ R = [ -\sum_{n=1, l=1}^{N, L} min(dist(a_n, l))] - c, \]  
where $N$ is the number of agents and $L$ is the number of landmarks in the environment. The function \textit{dist()} calculates the distance in terms of the agents' and landmarks' $(x_i, y_i)$ positions in the environment. The number of collisions, \textit{c}, among the agents and is set as a penalty of -1 for each time two agents collide. The action set is discrete, corresponding to moving in the four cardinal directions or remaining motionless. Each agent's observation includes the relative positions of other agents and landmarks within the frame. Note that the local observations do not convey the velocity (movement direction) of other agents. Consistent with past work, the initial positions of all the agents and the landmark locations are randomized at the start of each episode, and each episode ends after 25 time steps. 

To test \algo, we modify this task so that a centralized agent receives the union of local agents' observations at every time step. We also conduct relevant ablation studies on modified versions of this environment (exhibiting graceful degradation to the system, particularly to what the local agents can observe) to understand the contribution of \algo's central agent to the overall system. 
We are most interested in this environment because of the motivating robotic warehouse example. 

\subsection{Multi-Agent Walker} 
\begin{figure}[t]
	\centering
	\includegraphics[width=0.75\linewidth]{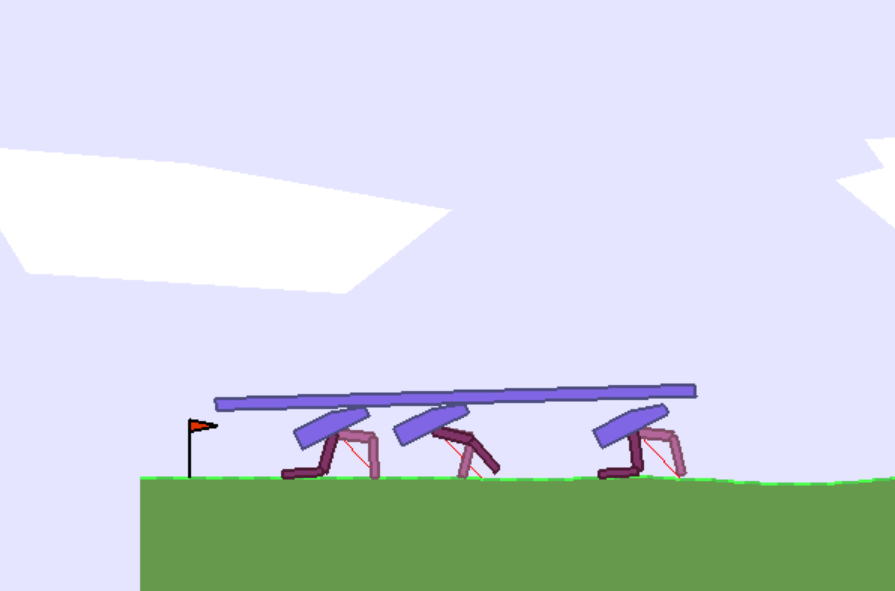}
	\caption{Multi-Agent Walker 
} 
	\label{fig:multiwalker}
	\Description{Test bed for the proposed architecture.}
\end{figure}

Multi-agent walker is a more complex, continuous control benchmark locomotion task~\cite{gupta2017cooperative}. A package is placed on top of $n$ pairs of robot legs which can be controlled. The agents must learn to move the package as far to the right as possible, without dropping it. The package is large enough (it stretches across all of the walkers) that the agents must cooperate. The environment demands high inter-agent coordination for them to successfully navigate a complex terrain while keeping the package balanced. This environment supports both team and individual rewards, but we focus on the latter, to demonstrate the effectiveness of \algo\ in individual feedback settings (as team rewards were already explored in the cooperative navigation task). Each walker is given a reward of -100 if it falls and all walkers receive a reward of -100 if the package falls. Throughout the episode, each walker is given an additional reward of +1 for every step taken. However, there is no supplemental reward for reaching the destination. By default, the episode ends if any walker falls, the package falls, after 500 time steps, or if the package successfully reaches the destination. Each agent receives 32 real-valued numbers, representing information about noisy Lidar measurements of the terrain and displacement information related to the neighbouring agents. The action space is a 4-dimensional continuous vector, representing the torques in the walker's two joints of both legs. Figure \ref{fig:multiwalker} is an illustration of the environment with $n=3$ agents. 

Similar to the Cooperative Navigation domain, this environment is tweaked so that all local agents in it can transmit their private observations to the central agent. This environment was chosen primarily to test our approach in a continuous action domain and in cases where individual agents receive their own local rewards instead of global team rewards. It is also significantly more challenging than the cooperative navigation domain. 

\subsection{Implementation Details} 


In both domains, two networks were used --- one for the central agent and the second for the independent agents. PPO was used to update the decision-making policy for the local agents. For the actor and critic networks of these individual agents, 2 fully-connected multi-layer perceptron layers were used to process the input layer and to produce the output from the hidden state. Three different variants were use to train the central agent. The first variant, $\algo$v1, had its own actor-critic network and learned its policy to communicate using PPO. In this case, the central agent's rewards were the same as those received by the local agents from the environment. In the second variant, $\algo$v2, the central agent used a multi-layer perceptron to output real-valued messages for the local agents and learned directly via the gradients passed back to it from the local agent network using backpropagation. In the final variant, $\algo$v3, instead of directly passing the messages to the local agent, we pre-processed the messages using a regularisation unit, RU (as described earlier in Section \ref{approach}). The central agent produces a vector of floating points scaled to $[-1, 1]$ and is consistent across domains. A tanh activation function is used everywhere except for the local agent's output. Local agents in cooperative navigation use a soft-max over the output, corresponding to the probability of executing each of the five discrete actions. Local agents in the multi-agent walker task use 4-output nodes, each of which model a multivariate Gaussian distribution over torques. 

Trials were run for 500,000 episodes in the cooperative navigation and 35,000 in the multi-agent walker environments. The training times of baselines and \algo\ were similar --- ~12 hours and ~14 hours, respectively, for 500,000 episodes on cooperative navigation and ~14 hours and ~18 hours, respectively, for 35,000 episodes in multi-agent walker. We set $\gamma = 0.95$ for all experiments. After having tried multiple learning rates $\{0.01, 0.001, 0.002, 0.005, 3 \times 10^{-4}, 1 \times 10^{-2}, 3 \times 10^{-3}\}$ and training batch sizes in PPO --- we found $3 \times 10^{-3}$ and 800, respectively, to work best for both the centralized agent and the independent agents in both domains. Five independent trials with different random seeds were performed on both environments to establish statistical significance. The clip parameter for PPO was set to 0.2. Empirically, we found that a message length of 1 was enough to perform well in both the navigation and multi-agent walker tasks.

\section{Results} 
\label{results}
This section describes the experiments conducted to test \algo's potential of encapsulating and communicating learned messages and speeding up independent learning 
on the two environments --- cooperative navigation and multi-agent walker --- whose details are described in the previous section. All the curves are averaged over five independent trials. Additional ablative studies were performed on the modified versions of cooperative navigation environment.

\subsection{Cooperative Navigation Results} 

We investigated in detail the learning of independent learners in the cooperative navigation environment under different situations. Learning curves used for evaluation are plots of the average reward per local agent as a function of episodes. Moreover, to smooth the curves for readability, a moving average with a window size of 1500 episodes was used for each of the cases. 

\begin{figure}[t]
	\includegraphics[width=\linewidth]{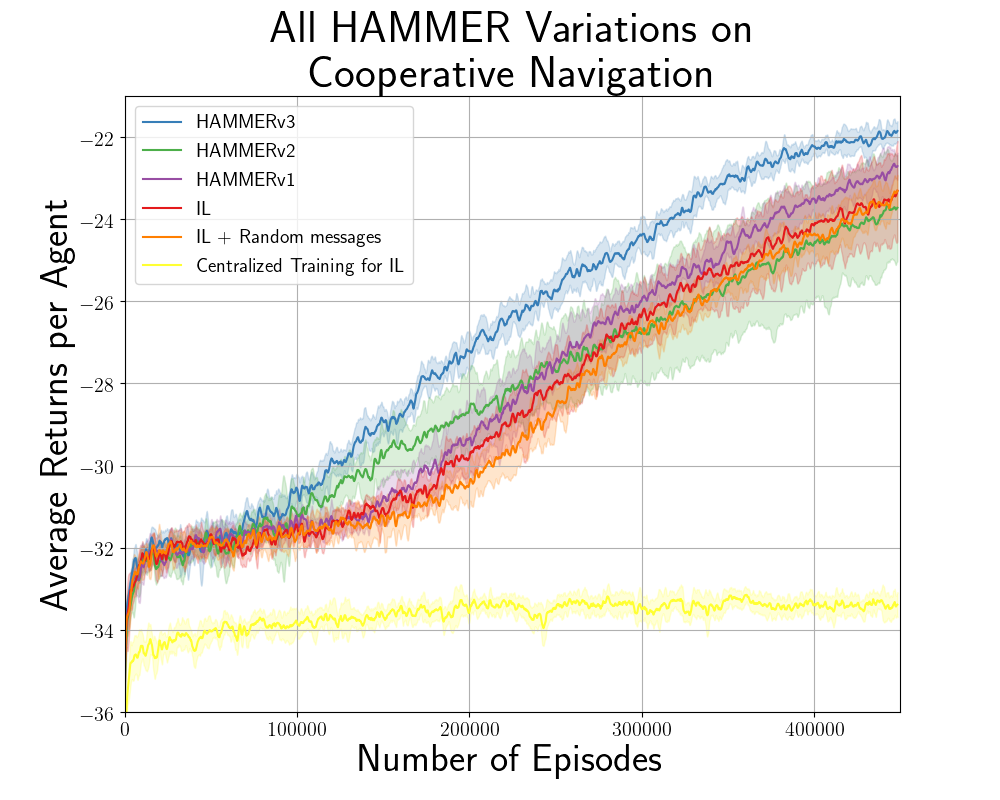} 
	\caption{\algo\ agents outperform independent learners and a centrally learned policy for independent agents. Providing the local agents with random messages causes degraded performance (as expected).} 
	\label{fig:random} 
	\label{fig:centralized} 
	\label{fig:hammercn} 
\end{figure}

At first, we let the local agents learn independently, without any aid from other sources. The corresponding curve (red), as shown in Figure \ref{fig:hammercn}, can also act as our baseline to evaluate learning curves obtained when the learners are equipped with additional messages. As described earlier, the experimental setup is consistent with earlier work \cite{gupta2017cooperative}. 


Next, we used a central agent to learn messages, as described by \algo,  and communicated them to the local learners to see if the learning improved. As described earlier (Section \ref{approach}) \algo's central agent is trained using more than one strategies and we compare the results here. First, the central agent learned to communicate by employing PPO, and used the local agents' team-based reward as its feedback ($\algo$v1). $\algo$v1 performed only slightly better than independent learners alone, over a training period of 500,000 episodes, as can be seen in Figure \ref{fig:hammercn} (purple). This training strategy does provide a small amount of improvement, which could be because of two types of problems: \textit{(Case 1)} The central agent may emit a relevant message to a local agent, but the local agent is unable to intercept it correctly resulting in a poor reward. In such a case, the central agent gets penalized, even though it performed its part well.  \textit{(Case 2)} The central agent emits relevant messages to some local agents, and non-relevant messages to others. In such a case, the central agent is penalized in spite of emitting relevant messages for some agents. 

Now, to avoid the problems encountered in the previous case, gradients from the local agent's network were pushed to \algo's central agent network, by directly connecting the outputted communication vectors to the input of the local agent's network ($\algo$v2). Letting gradients flow in this manner, gives the central agent richer feedback, thus reducing the required amount of learning by trial and error, and easing the discovery of effective real-valued messages. Since these messages function as any other network activation, gradients can be passed back to the central agent's network, allowing end-to-end backpropagation across the entire framework. As shown in Figure \ref{fig:hammercn}, $\algo$v2 (green) performs significantly better than unaided independent learners as well as $\algo$v1. However, \algo\ agents seem to slow after 250,000 episodes. We speculate that a larger action space (real-valued messages) causes the central agent learns to ``over-encode" the global information available to it. Having included this extra information along with the relevant pieces would have masked the private observations of the local agents, slowing their progress. 


Finally, addressing the problem of ``over-encoding" faced in the previous case, the emitted messages were first processed by a regularisation unit --- RU($m_i$) = Logistic($\mathbf{N}$ ($m_i$, $\sigma$)), to encourage discretization of the messages (by pushing the activations of the communication vectors into two different modes during training, i.e., where the noise is minimized), and then passed to the local agent's network ($\algo$v3). Figure \ref{fig:hammercn} illustrates that $\algo$v3 (blue) outperforms all the other cases. Hence, using a regularization unit to limit the central agent's action space was essential for learning a better communication policy. Using a noisy channel in this manner has been 
useful in other relevant works too --- differentiable inter-agent learning~\cite{NIPS2016_6042}, training document models~\cite{hinton2011discovering} and performing classification~\cite{courbariaux2016binarized}. For our experiments, we used a value of $\sigma=0.2$. 

From these experiments, our results show that: (1) \algo\ agents were able to learn much faster when compared to independent local agents, and (2) the central agent was able to successfully learn to produce useful smaller messages. Recall that the total global observation vector has $18 \times 3=54$ real-valued numbers (for 3 agents and 3 landmarks in the environment), and our message uses only 1. 

To help evaluate the communication quality, random messages of the same length were generated and communicated to the local agents to see if the central agent was indeed learning relevant or useful messages. As expected, random messages induce a larger observation space with meaningless values and degrade the performance of independent learners (Figure \ref{fig:random}, orange curve). This also supports the claim that the central agent is learning much better messages to communicate (rather than sending random values) as it outperforms the independent learning of agents provided with random messages. Note here that the local agents are free to learn to ignore unhelpful messages and independent agents' learning while receiving random messages only degrades performance slightly. 

To confirm that the central agent is not simply forwarding a compressed version of the global information vector to all the agents, we let the independent agents learn in a fully centralized manner, using a joint observation space. The performance drops drastically in this case (Figure \ref{fig:centralized}, yellow curve). This suggests that the central agent in  \algo\ is learning to encapsulate only partial but relevant information as messages and communicates them to facilitate the learning of local agents. Complete information would have become too overwhelming for the localized agents to learn how to act and hence slowed the progress, as is clear by the curve shown in Figure \ref{fig:centralized}. It has been shown previously that this approach does not perform well in domains such as pursuit, water world and multi-agent walker \cite{gupta2017cooperative}. We confirm the same in the cooperative navigation environment for n=3 agents. 

\begin{figure}[t]

	\includegraphics[width=\linewidth]{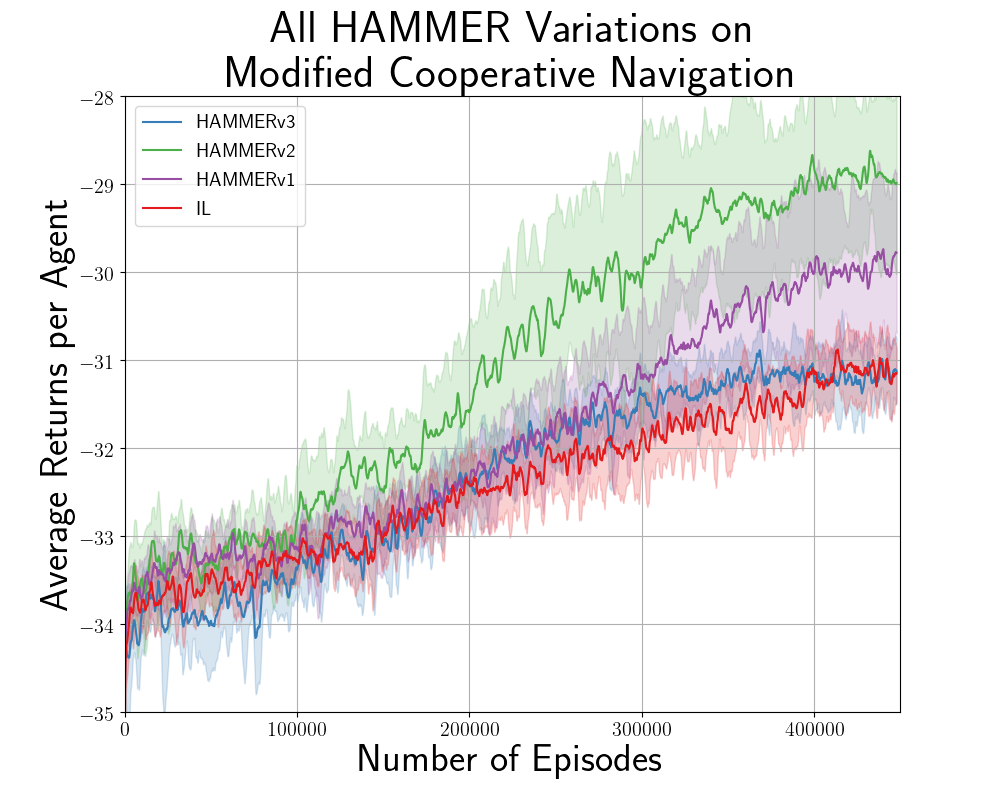} 
	\caption{This modified cooperative navigation environment prevents the local agents from observing each other, necessitating communication. \algo's ability to perform in this setting shows that the central agent is indeed learning effective communication to help the local agents coordinate. } 
	\label{fig:modifiedcn} 
\end{figure}

\textbf{Ablation Studies.} First, we perform an ablation experiment to validate that \algo's central agent is indeed facilitating independent agents' learning and helping them coordinate by sending out relevant messages. For this, we modify the environment preventing agents from seeing each other. Local agents be unable to observe other agents and would have to rely on the central agent for coordinating and covering respective landmarks. Results of \algo\ in this scenario, compared to unaided independent learners, are in Figure \ref{fig:modifiedcn}. As expected, it became difficult for the independent learners to learn the cooperation-intensive task on their own. On the other hand, \algo\ agents learned substantially better policies faster. Note here that $\algo$v2 in this case performs better than $\algo$v3. We speculate that this is due to local agents' greater dependency on central agent's communicated messages for coordinating, and hence, \algo\ requiring a larger spectrum to encode and communicate more information into its messages. 

\begin{figure}[t]
	\includegraphics[width=\linewidth]{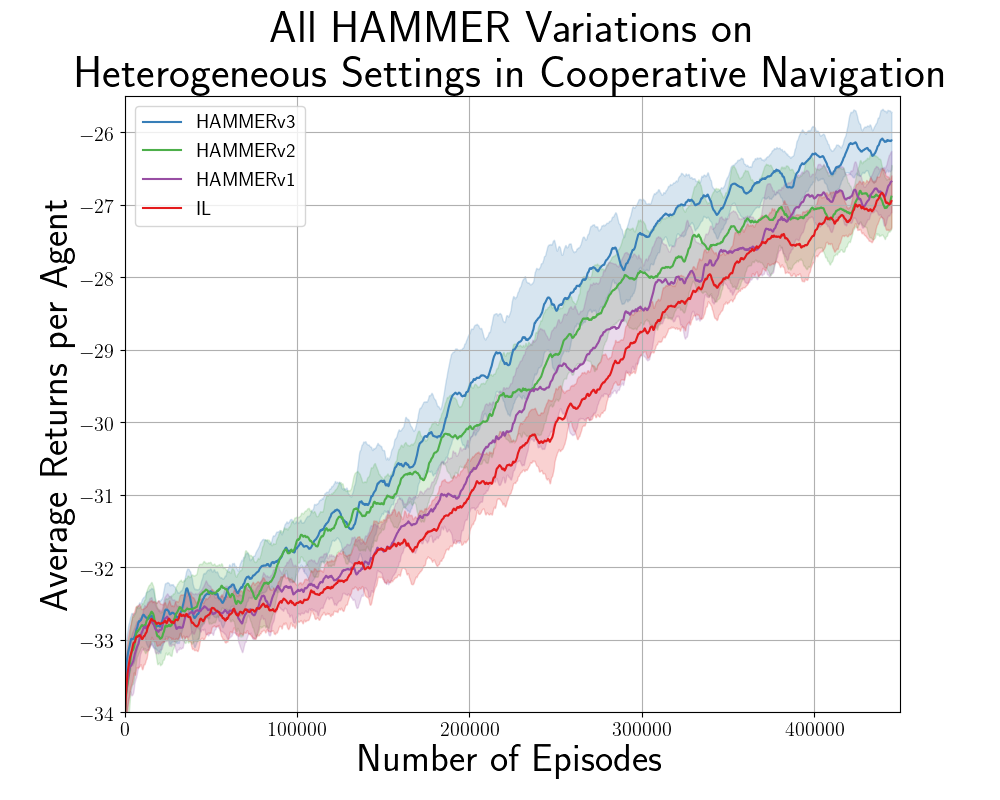} 
	\caption{This version of cooperative navigation has heterogeneous local agents --- one was unable to observe the other agents, while the other two were unmodified. \algo\ is again successful.} 
	\label{fig:heterogeneitycn} 
\end{figure}

In our second ablation study, we validate whether \algo\ would generalize to a setting with heterogeneous local agents. This experiment uses an environment where one of the local agents was unable to observe other agents, while the other two agents still received their original observations. All local agents still learn and use the same policy, but one of them cannot see the others. As expected, independent learners had increased difficulty in learning to coordinate (Figure \ref{fig:heterogeneitycn}), and were outperformed by all variants of learning strategies for \algo. We conclude that \algo\ generalizes to heterogeneous settings too. 

In summary, \algo\ successfully summarizes relevant global knowledge into small real-valued messages to individual learners that outperforms other cases --- (1) unaided independent learning, (2) independent agents supplied with random messages, and (3) fully centralized training of independent agents. Specifically, $\algo$v3 (i.e., \algo\ trained by directly passing message gradients from the local agent's network to the central agent using backpropagation, and its messages preprocessed using a regularization unit) outperforms the other training strategies, learning a significantly better policy than local agents learning independently.


\subsection{Multi-Agent Walker Results} 
\begin{figure}[t]
	\includegraphics[width=\linewidth]{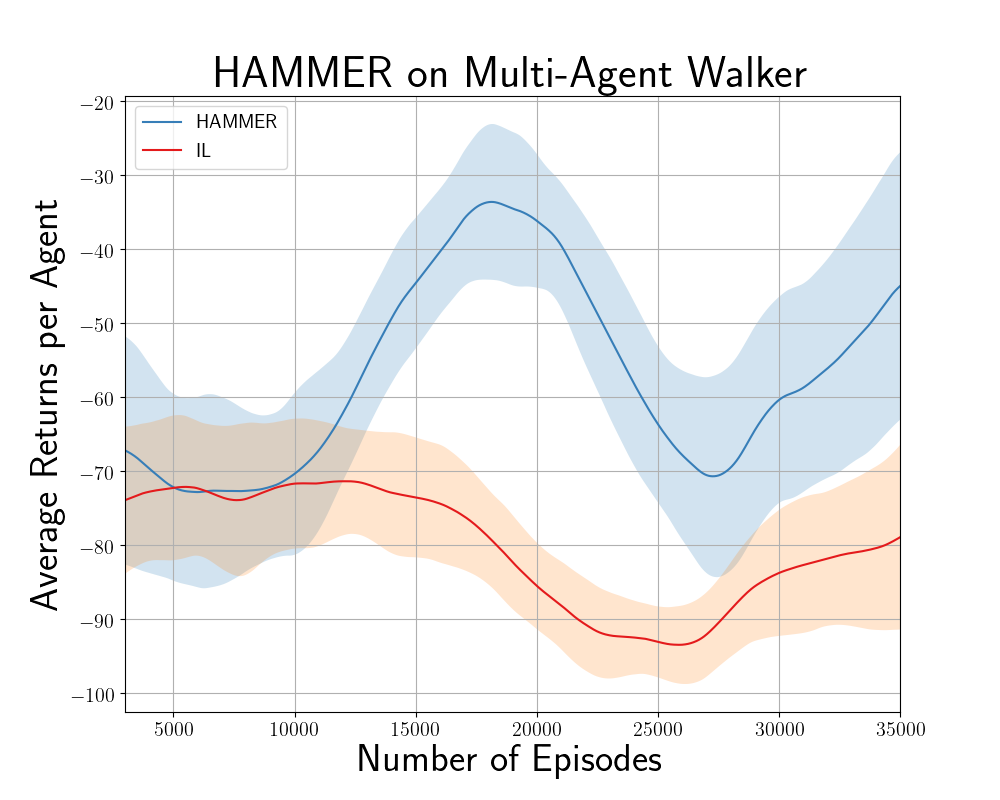} 
	\caption{\algo\ considerably improves the performance over independent learners in the multi-agent walker task too. } 
	\label{fig:mw} 
\end{figure}

This section shows that \algo\ also works in a multi-agent task with continuous control and individual rewards. Figure \ref{fig:mw} shows that $\algo$v3 agents perform considerably better than unaided independent local agents. Like in the cooperative navigation task, results here are averaged over five independent trials, with an additional 5000-episode moving window to increase readability. 
\algo\ performing better in a domain like Multi-Agent Walker confirms the generalization of the approach to continuous action spaces and different reward structures.\footnote{We did not include the results corresponding to the other baselines here as Gupta et al.\ \cite{gupta2017cooperative} already show that independent learners outperform a centralized policy in the multi-agent walker domain.} 

The results for the two test domains show that (1) heterogeneous agents successfully learn messaging and enable multi-level coordination among the independent agents to enhance reinforcement learning using \algo\ approach, 
(2) \algo\ generalizes to heterogeneous local agents in the environment, 
(3) the approach works well in both discrete and continuous action spaces, and (4) the approach performed well with both individual rewards and global team rewards.

\section{Conclusion} 
\label{conclusion}
This paper presented \algo, an approach used to achieve multi-level coordination among heterogeneous agents by learning useful messages from a global view of the environment or the multi-agent system. Using \algo, we addressed challenges like non-stationarity and inefficient coordination among agents in MARL by introducing a powerful all-seeing central agent in addition to the independent learners in the environment. Our results in two domains, cooperative navigation and multi-agent walker, showed that \algo\ can be generalized to discrete and continuous action spaces with both global team rewards and localized personal rewards. \algo\ also proved to generalize to heterogeneous local agents in the environment 
We believe that the key reasons for the success of \algo\ were two-fold. First, we leveraged additional global information like global states, actions and rewards in the system. Second, centralizing observations has its own benefits, including helping to bypass problems like non-stationarity in multi-agent systems and avoiding getting stuck in local optima~\cite{whiteson2011protecting, johanson2011accelerating}. Several works, related to ours (discussed earlier) demand powerful agents in the environment --- ones that can transmit to other agents and/or are capable of modeling other agents in the system. This might not always be feasible, thus motivating \algo, where only one central agent needs to be powerful while the independent learners can be simple agents interacting with the environment based on their private local observations augmented with learned messages. Warehouse management and traffic lights management are two example applications that could fit these assumptions well.


There are several directions for future work from here. Both the domains used in this work involved low-dimensional observation spaces and seem to offer substantial global coverage on combining local observations. \algo's results in multi-agent settings with tighter coupling and further complex interactions involved among agents such as in autonomous driving in SMARTS \cite{zhou2020smarts} or heterogeneous multi-agent battles in StarCraft \cite{peng2017multiagent} could help better appreciate the significance of the method. Additionally, more complex hierarchies could be used, such as by making several central agents available in the system. In this work, we performed an initial analysis of message vectors communicated by \algo, but additional work remains to better understand if and how \algo\ 
tailors messages for the local agents using its global observation. It would also be interesting to further study how RL learns to encode information in messages and to understand what the encoding means. It would also be interesting to test the scalability of \algo\ to a larger number of local agents in the environment. 
Lastly, in our setting, communication is free --- future work could consider the case where it was costly and attempt to trade off the number of messages sent with the learning speed of \algo. If the central agent had the ability to communicate small amounts of data occasionally, would it still be able to provide a significant improvement? 



\vspace{20pt}
\begin{acks} 
This work commenced at Ericsson Research Lab Bangalore, and most of the follow-up work was done at the International Institute of Information Technology - Bangalore.\footnote{As part of Nikunj Gupta's Master's Thesis titled ``Fully Cooperative Multi-Agent Reinforcement Learning".} Part of this work has taken place in the Intelligent Robot Learning (IRL) Lab at the University of Alberta, which is supported in part by research grants from the Alberta Machine Intelligence Institute (Amii), CIFAR, a Canada CIFAR AI Chair, and NSERC. 
We would like to thank Laura Petrich, Shahil Mawjee and anonymous reviewers for comments and suggestions on this paper.
\end{acks}

\clearpage 

\bibliographystyle{plain} 
\bibliography{rl}


\end{document}